\documentclass[journal,twoside,web]{ieeecolor}
\usepackage{generic}
\usepackage{hyperref}
\usepackage{cite}
\usepackage{amsmath,amssymb,amsfonts}
\usepackage{algorithmic}
\usepackage{graphicx}
\usepackage{textcomp}
\usepackage{textgreek}
\newcommand{\txc}{}%{\textcolor{red}}
\def\BibTeX{{\rm B\kern-.05em{\sc i\kern-.025em b}\kern-.08em
    T\kern-.1667em\lower.7ex\hbox{E}\kern-.125emX}}
\begin{document}

\bstctlcite{IEEEexample:BSTcontrol}
\title{Integration of 3-level MoS$_2$ multi-bridge channel FET with 2D layered contact and gate dielectric}
\author{Hitesh S, Pushkar Dasika, Kenji Watanabe, Takashi Taniguchi, and Kausik Majumdar
\thanks{Hitesh S, Pushkar Dasika and Kausik Majumdar are with Department
of Electrical Communication Engineering, Indian Institute of Science, Bengaluru 560012, India (e-mail: kausikm@iisc.ac.in). }
\thanks{Kenji Watanabe is with the Research Center for Functional Materials, National Institute for Materials Science, 1-1 Namiki, Tsukuba 305-0044, Japan}
\thanks{Takashi Taniguchi is with the International Center for Materials Nanoarchitectonics, National Institute for Materials Science,  1-1 Namiki, Tsukuba 305-0044, Japan.}
}
\maketitle
\begin{abstract}
 Multi-bridge channel field effect transistor (MBCFET) provides several advantages over FinFET technology and is an attractive solution for sub-5 nm technology nodes. MBCFET is a natural choice for devices that use semiconducting layered materials (such as, MoS$_2$) as the channel due to their dangling-bond-free ultra-thin nature and the possibility of layer-by-layer transfer. MoS$_2$-based MBCFET is thus an attractive proposition for drive current boost without compromising on the electrostatics and footprint. Here we demonstrate a 3-level MoS$_2$ MBCFET, where each vertically stacked channel is dual-gated to achieve a saturation current of 174.9 $\mu$A (which translates to \txc{90 $\mu$A per $\mu$m footprint width (@2.7 $\mu$m channel length), a near-ideal sub-threshold slope of 63 mV/dec, and an on-off ratio $>$$10^8$. This work sets the benchmark for layer-material-based MBCFET in terms of the number of parallel channels integrated, simultaneously providing high drive current and excellent electrostatic control.}
\end{abstract}

\section{Introduction}

\begin{figure}[h!]
\includegraphics[width=\columnwidth]{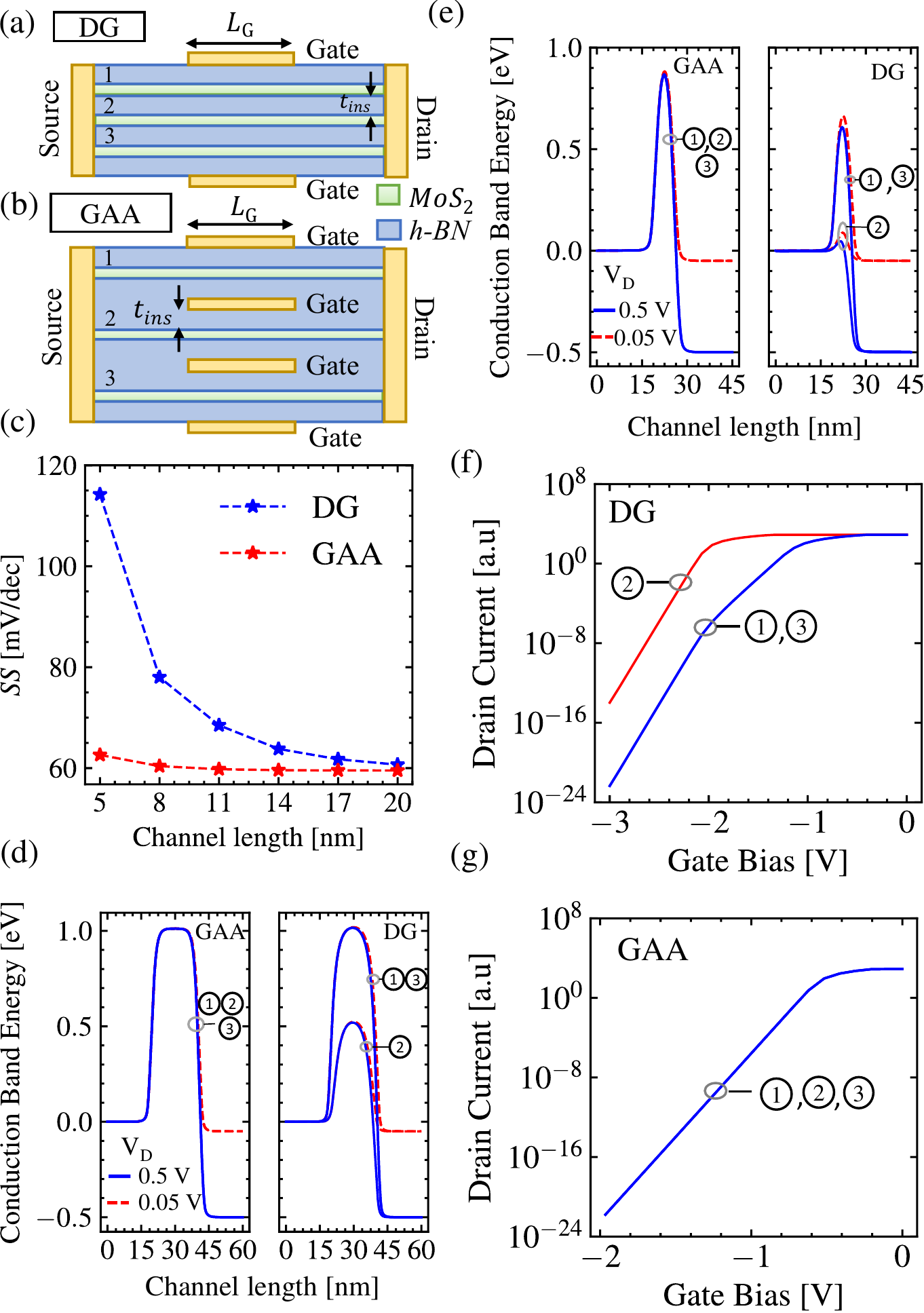}
\caption{Schematic representation of (a) double gate (DG) and (b) gate-all-around (GAA) structure. (c) Simulated sub-threshold slope as a function of channel length for both structures.  (d-e) Simulated conduction band profiles of GAA (left panels) and DG (right panels) along the three channels (marked 1, 2, and 3) with a channel length of (d) 20 nm  and (e) 5 nm. Simulated transfer characteristics of (f) DG and (g) GAA structures under depletion-mode (normally-on) operation. A threshold shift is found in the central channel of DG structure. We assumed an effective oxide thickness (EOT) of of 2 nm for each of the gate insulators.}
\end{figure}

The transistor dimensions have shrunk aggressively over the past few decades, helping to improve performance and pack more functionality per unit area. During the scaling pathway, non-planar FinFET technology replaced the planar technology to overcome short channel effects.
However, for beyond-5 nm technology nodes, there are several challenges posed by the FinFET technology. This includes continued fin width scaling to control short channel effects, fin height scaling to enhance drivability, and lack of flexibility in enhancing effective width through a discrete number of fins.  MBCFET \cite{bae20183nm,barraud20207,lee2004novel,huang2020high,yoon2004sub, ritzenthaler2018vertically,mertens2016vertically} is a viable solution to support the continuously increasing demand for enhanced effective channel width without compromising the device footprint. This also brings back the flexibility of the planar technology in terms of continuous width scaling.

However, below a certain thickness, the mobility in Si nano-sheets decreases drastically due to surface roughness. Therefore the ultimate utilization of MBCFET architecture would require an atomically thin sheet of semiconductor with excellent electrostatic gate control and clean interfaces. This makes two-dimensional channel materials (such as MX$_2$, M = Mo, W; X = S, Se) an exceptional choice \cite{huang2021ultrathin,liu2021two,xia2022wafer,zhou2018three,xiong2021demonstration}. The lack of dangling bonds allows these materials to be scaled down to a monolayer-thickness ($\sim 0.7$ nm) without compromising on carrier mobility. The excellent electrostatics of such ultra-thin sheets would thus allow superior scalability \cite{majumdar2014scalability}.

Here we demonstrate a van der Waals (vdW) heterostructure based MBCFET, consisting of three vertically stacked 1L-MoS$_2$ channels, each being dual gated with an hBN/graphene gate stack \txc{and contacted by graphene layers. Our best device shows a  saturation current of 90 $\mu$A per $\mu$m footprint width (@ effective channel length of 2.7 $\mu$m) while maintaining near-ideal subthreshold slope and an on-off ratio $>$$10^8$}.
\section{Proposed device and scalability projection}

The scalability of a transistor to shorter channel lengths is an essential aspect when selecting a suitable device design. To illustrate the electrostatic control at scaled geometry, we perform a 2D simulation of two different MBCFET architectures. In one of the designs, a pair of double-gates control all the three channels of the transistor [see Fig. 1(a)], which we call a double-gated (DG) structure. In the other design, each of the three  channels is controlled by dual-gates [Fig. 1(b)], which we call a gate-all-around (GAA) structure.

The simulation is performed by self-consistently solving the 2D Poisson and ballistic transport equations in the device \cite{rahman2003theory}.
The predicted sub-threshold slope (SS) of each design is shown in Fig. 1(c), as a function of the channel length. We observe that although the DG structure offers an SS of $\sim$ 60 mV/dec at longer channel lengths ($\geq$ 20 nm), the SS of the device increases to $\sim$ 114 mV/dec at a gate length of 5 nm. In contrast, the SS increase in the GAA device is minimal, and even at a channel length of 5 nm, the device exhibits an SS of 62 mV/dec.

The conduction band profiles along different channels (marked as 1, 2, and 3) in the GAA and DG structures are shown in Figs. 1(d) and 1(e) with channel lengths of 20 and 5 nm, respectively. Even at a channel length of 20 nm, the DG structure starts displaying drain-induced barrier lowering (DIBL). However, the GAA structure shows no such sign. The simulated band profile of all the channels in a GAA structure is essentially the same, but in the DG structure, the barrier height of the central channel is significantly lower.

The lack of electrostatic control on the central channel in the DG structure is further illustrated by the simulated transfer characteristics in Fig. 1(f). This would cause different threshold voltages for different channels, which is undesirable. However, the symmetry of the GAA structure mitigates this problem efficiently [Fig. 1(g)]. This scalability of the GAA structure prompts us to fabricate this design, although this structure has a more involved fabrication procedure.

\section{Device Fabrication and measurement}

\begin{figure}[h!]
\includegraphics[width=\columnwidth]{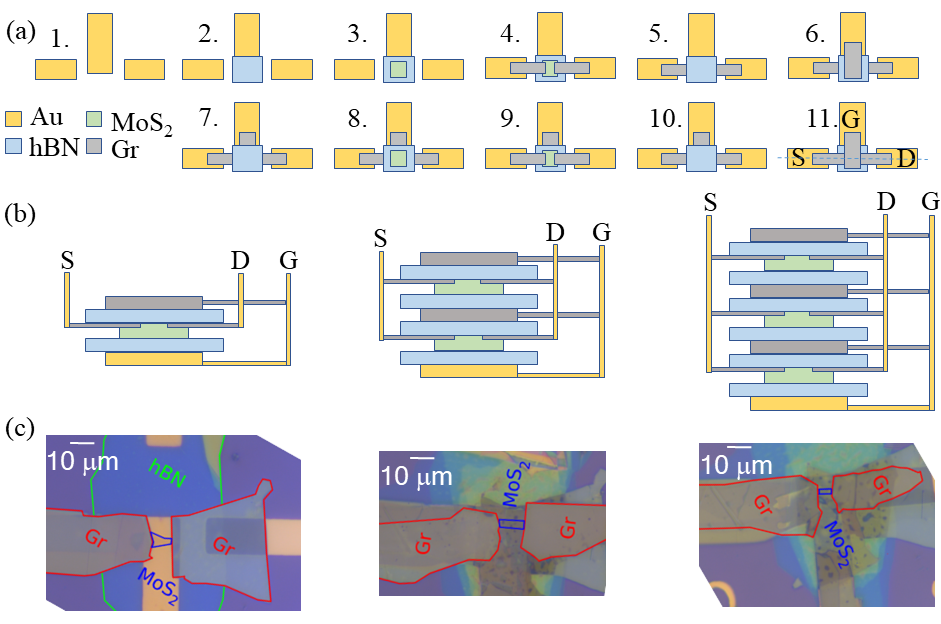}
\caption{(a) Step-by-Step process showing the fabrication methodology for a three-channel stack. (b) Cross-section view when cut along the drain-source electrodes [dashed line in 11 of (a)] and (c) optical images at the one-, two-, and three-channel configurations \txc{(device A)}. Scale bar is 10 $\mu$m.}
\end{figure}

\begin{figure}[h!]
\includegraphics[width=\columnwidth]{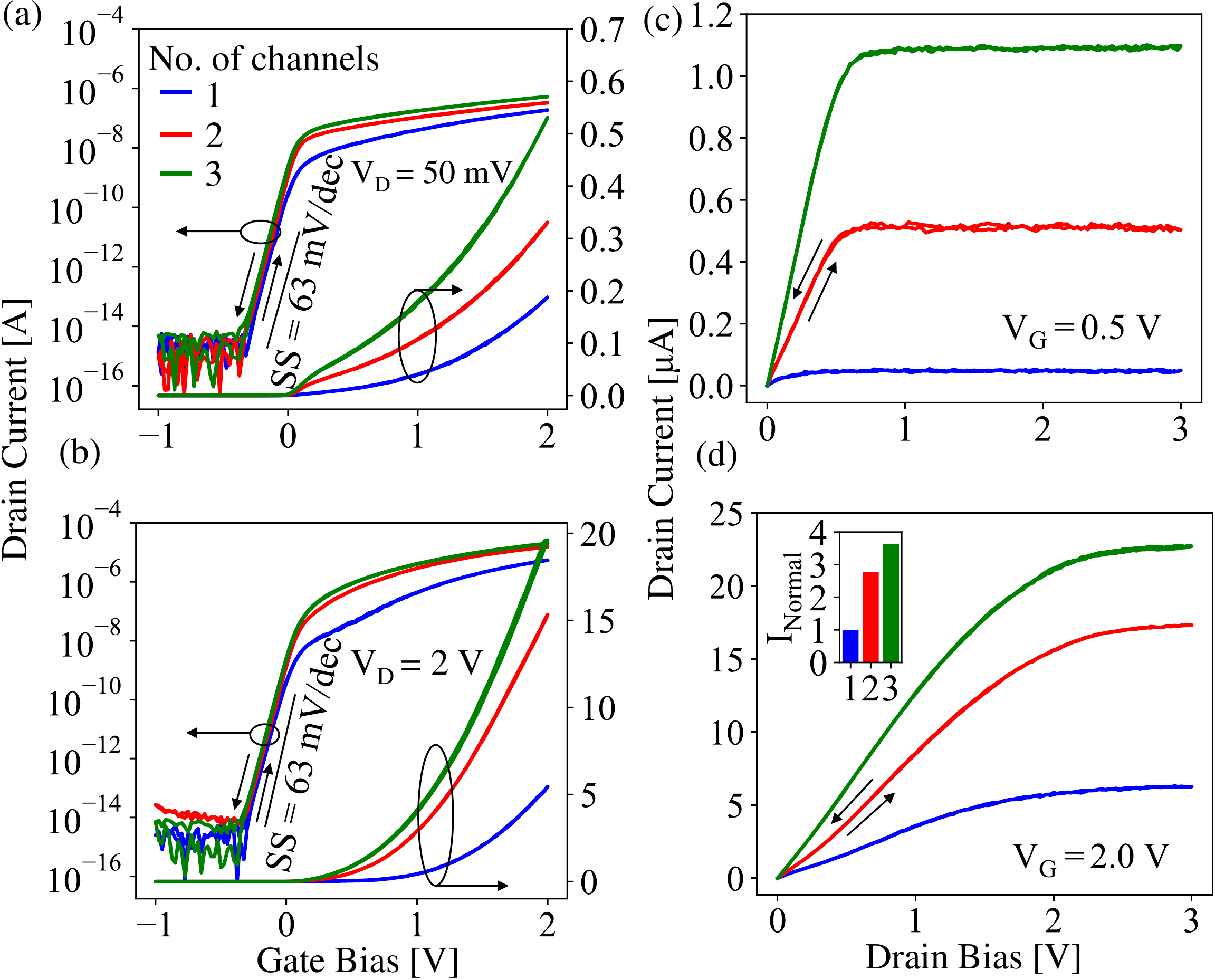}
\caption{(a-b) Transfer characteristics of device A (device parameters in Table \ref{tab:1}) with number of channels at (a) $V_D=0.05$ V and (b) $V_D=2$ V. Output characteristics with number of channels at (c) $V_G=0.5$ V and (d) $V_G=2$ V. Inset: Scaling of $I_D$ with number of channels at $V_G=2$ V. The forward and reverse sweeps are shown by black arrows.}
\end{figure}

The step-by-step fabrication process of the MBCFET is schematically illustrated in Fig. 2(a). The fabrication starts with creating Au pads on 285 nm SiO$_2$/Si wafer using optical lithography, metal deposition, and subsequent lift-off.  Mechanically exfoliated few-layer-thick hBN is then dry-transferred (using PDMS) onto a pre-patterned Au electrode. An exfoliated 1L-MoS$_2$ flake is then transferred onto the hBN to form the first channel. Few-layer graphene (FLG) flakes with sharp edges are used to make contact from the 1L-MoS$_2$ to the Au electrodes, which form the drain and source contacts. This stack is then capped with another few-layer-thick hBN. The entire stack is annealed in vacuum (pressure $\sim 10^{-6}$ torr) at $200^o$C for 3 hours to remove trapped air/residues between the layers. Another FLG flake is then transferred on top to form the double gate structure. This FLG layer acts both as the top gate for the first channel and also as the bottom gate of the second channel. The entire stack at this point is schematically illustrated in the left panel of Fig. 2(b). The stack is further built by repeating the above steps to make the two-channel and three-channel device.

The thickness of the hBN gate dielectric is chosen to be $\sim 15$ nm for all the channels. The cross-section of the device and the corresponding optical image at the three different stages are shown in Figs. 2(b) and (c), respectively. A combination of the annealing steps and the atomic smoothness of the thin FLG S/D contact, hBN gate dielectric and FLG gate contact is likely to help maintain the flatness of the channels \cite{Cheng20222Dinterfaces}.

Measurements are taken at each of these three stages to obtain the performance of one-, two-, and three-channel device. All the three source electrodes (and similarly, the three drain electrodes and the four gate electrodes) are shorted together in the device. Two such devices (devices A and B) are fabricated and measured. All measurements are performed at 295 K using a Keithley 4200 parametric analyzer, keeping the device in Lakeshore CRX-6.5K probe station under vacuum (pressure $<10^{-4}$ torr).

\section{Results and Performance}
The effective channel length and width for the two fabricated devices are shown in Table \ref{tab:1}. The measured transfer characteristics of device A are depicted in Fig. 3(a-b), both in log (left axis) and linear (right axis) scale for two different drain biases ($V_D$), namely 0.05 and 2 V. Each plot contains the characteristics of the one-, two-, and three-channel configurations. The off current ($I_{off}$) of the 3-channel configuration is around $\sim 10$ fA (which is the limit of our measurement accuracy) at $V_D=50$ mV, which increases to $\sim 100$ fA at $V_D=4$ V.  This leads to a large on-off ratio ($I_{on}/I_{off}$)  of $>10^8$ indicating the robustness of the device to suppress the leakage current.  The SS obtained from Fig. 3(a-b) is near-ideal, $\sim$ 63 mV/dec over several decades of drain current ($I_D$), and does not vary with $V_D$. This is attributed to the exceptional electrostatics due to the atomically thin monolayer MoS$_2$ channels and the clean  hBN/MoS$_2$ interfaces.

The output characteristics of the device are plotted in Fig. 3(c-d) for $V_G=0.5$ and $2$ V. We observe linear characteristics at lower $V_D$ indicating ohmic behaviour of the FLG-MoS$_2$ contacts. Due to the vdW nature of the contact interface, we expect a highly de-pinned contact \cite{murali2021accurate, roy2014field}. The good quality of the contact further helps to achieve excellent drain current saturation at higher $V_D$, as depicted by the characteristics.

The scaling of the on current with an increase in the number of stacked channels is clearly illustrated in Fig. 3(c-d),  which is an efficient way of effective-width scaling without consuming an additional horizontal footprint.

From Fig. 3(a-b), we note that the threshold voltage ($V_{th}$) of the three different channel configurations exhibits a small variation with the number of channels stacked. This arises due to a variation in the thickness of the gate dielectric and the contact quality between FLG and MoS$_2$. \txc{The use of Au (having higher work function compared with FLG) back gate electrode for the first channel may also contribute to the same.} As a result of such $V_{th}$ variation with the number of channels, the effective enhancement in the current with the number of channels remains a function of the gate voltage ($V_G$). The inset of Fig. 3(d) shows the scaling of the on-current at $V_G=2$ V, after normalizing with respect to the channel dimensions.

\txc{Note that all the characteristics in Fig. 3 are hysteresis-free (sweep directions indicated by black arrows).} MoS$_2$ channel-based FET devices are often prone to hysteresis, which largely arises due to trapping from moisture, residues, and other surface effects. The vacuum annealing steps and all the channels being encapsulated by hBN/graphene from top and bottom effectively removed the hysteresis almost completely.

The transfer and output characteristics of the 3-channel configuration (device A) are shown in Fig. 4(a) for $V_G$ and $V_D$, both varying up to 4 V. The SS remains intact at $63$ mV/dec irrespective of the $V_D$, even up to 4 V. The performance of devices A and B is summarized in Table \ref{tab:1}. We obtain a saturation current of 174.9 $\mu$A  from device B under 3-channel configuration at $V_D=4$ V. \txc{This translates to a saturation current of 90 $\mu $A per $\mu$m footprint width, with an on-off ratio $>$$10^8$. The performance of devices A and B is compared \cite{Cheng2022HowTo} with other MoS$_2$ based MBCFETs in Table \ref{tab:2}.}

\begin{figure}[h!]
\includegraphics[width=\columnwidth]{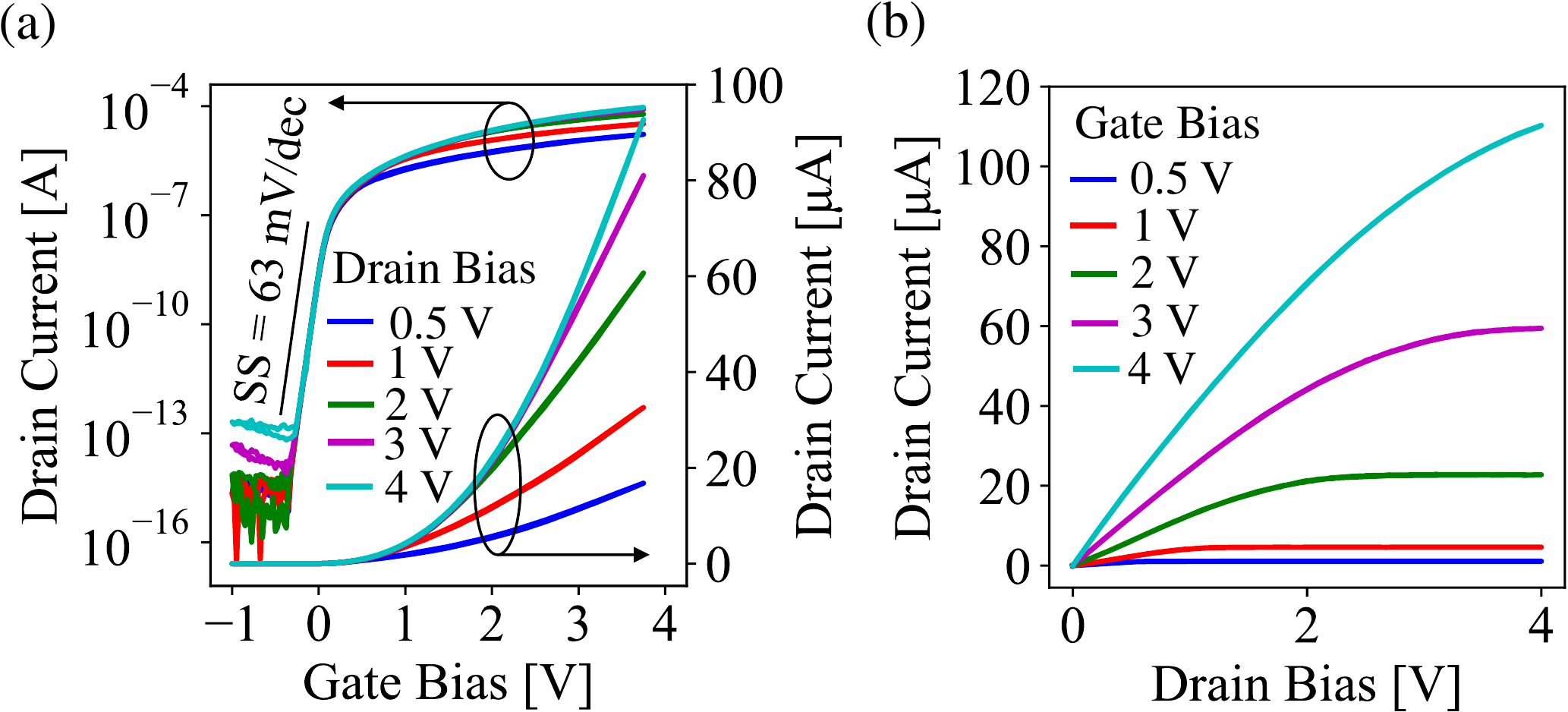}
\caption{(a) Transfer and (b) output characteristics of three-channel MBCFET (device A with parameters as in Table \ref{tab:1}) at high bias.}
\end{figure}

\section{Conclusion}
In conclusion, we demonstrated a 3-level MoS$_2$-channel MBCFET with a large drive current while maintaining a near-ideal sub-threshold slope and a large on-off ratio. The ability to perform a layer-by-layer transfer, coupled with the ultra-thin body of the channel shows a path towards integrating several more channels vertically - a viable option to boost the drive current without increasing the footprint. The demonstrated technique, when combined with a more optimized contact process, can further enhance the drive current. This is a promising solution not only for layered material-based highly scaled CMOS technology, but also for low-cost thin-film transistor (TFT) technologies.

\begin{table}[h!]
\centering
\caption{Summary of device parameters and performance\label{tab:1}}
\begin{tabular}{ |c|c|c| }
\hline
Parameter & Device A & Device B \\
\hline\hline
$L_{ch}$ (ch1, ch2, ch3) [$\mu$m] & 5.5, 4.5, 6.6 & 1.6, 4.5, 3.9 \\
\hline
$W$ (ch1, ch2, ch3) & 2.5, 1.6, 2.5 & 2.6, 1, 2.2 \\
\hline
Gate dielectric (hBN) thickness [nm] & $\sim$15 & $\sim$15\\
\hline
FLG S/D thickness [nm] & 10-15 & 10-15\\
\hline
On current (V$_G$=4V, V$_D$=4V) [\textmu A] & 110.28 & 174.9 \\
\hline
Off current [pA] & $<1$ & $<1$ \\
\hline
I$_{on}$/I$_{off}$ & $>10^8$ & $>10^8$\\
\hline
SS [mV/dec] & $63$ & $65$ \\
%\hline
%On current per footprint width & 295.9\textmu A/\textmu m.\textmu m & 419.1\textmu A/\textmu m.\textmu m \\
\hline
\end{tabular}
\end{table}
\vspace{-0.18in}
\begin{table}[h!]
\centering
\caption{Benchmarking with MoS\textsubscript{2} MBCFETs\label{tab:2}}
\begin{tabular}{|c|c|c|c|c|c|}
\hline
 Parameter & {This work (}A{)} & {This work (}B{)} & \cite{huang2021ultrathin} & \cite{zhou2018three} & \cite{xiong2021demonstration}\tabularnewline
\hline
\hline
$L_{ch}$ {[}$\mu m${]} & 5.4$^*$ & 2.7$^*$ & 12 & 0.37 & 0.16\tabularnewline
\hline
Channels & 3 & 3 & 2 & 2 & 2\tabularnewline
\hline
$I_{sat}$ {[}$\mu A$/$\mu$m{]} & 50 & 90 & 2.9 & 535 & 700\tabularnewline
\hline
$I_{on}/I_{off}$ & $>10^{8}$ & $>10^{8}$ & $10^{8}$ & $10^{6}$ & $10^{7}$\tabularnewline
\hline
SS {[}mV/dec{]} & 63 & 65 & 60 & 130 & 126\tabularnewline
\hline
\end{tabular}
$^*$$L_{ch}$ calculated as the harmonic mean of the lengths of the three channels
\end{table}

\bibliographystyle{IEEEtran} % We choose the "plain" reference style
\bibliography{IEEEabrv,references} % Entries are in the refs.bib file

\begin{thebibliography}{10}
\providecommand{\url}[1]{#1}
\csname url@samestyle\endcsname
\providecommand{\newblock}{\relax}
\providecommand{\bibinfo}[2]{#2}
\providecommand{\BIBentrySTDinterwordspacing}{\spaceskip=0pt\relax}
\providecommand{\BIBentryALTinterwordstretchfactor}{4}
\providecommand{\BIBentryALTinterwordspacing}{\spaceskip=\fontdimen2\font plus
\BIBentryALTinterwordstretchfactor\fontdimen3\font minus
  \fontdimen4\font\relax}
\providecommand{\BIBforeignlanguage}[2]{{%
\expandafter\ifx\csname l@#1\endcsname\relax
\typeout{** WARNING: IEEEtran.bst: No hyphenation pattern has been}%
\typeout{** loaded for the language `#1'. Using the pattern for}%
\typeout{** the default language instead.}%
\else
\language=\csname l@#1\endcsname
\fi
#2}}
\providecommand{\BIBdecl}{\relax}
\BIBdecl

\bibitem{bae20183nm}
G.~Bae, D.-I. Bae, M.~Kang, S.~Hwang, S.~Kim, B.~Seo, T.~Kwon, T.~Lee, C.~Moon,
  Y.~Choi, K.~Oikawa, S.~Masuoka, K.~Chun, S.~Park, H.~Shin, J.~Kim,
  K.~Bhuwalka, D.~Kim, W.~Kim, J.~Yoo, H.~Jeon, M.~Yang, S.-J. Chung, D.~Kim,
  B.~Ham, K.~Park, W.~Kim, S.~Park, G.~Song, Y.~Kim, M.~Kang, K.~Hwang, C.-H.
  Park, J.-H. Lee, D.-W. Kim, S.-M. Jung, and H.~Kang, ``3nm {{GAA}} technology
  featuring {{Multi-Bridge-Channel FET}} for low power and high performance
  applications,'' in \emph{2018 IEEE International Electron Devices Meeting
  (IEDM)}, Dec. 2018, pp. 28.7.1--28.7.4, {DOI} :
  {\href{https://doi.org/10.1109/IEDM.2018.8614629}{10.1109/IEDM.2018.8614629}}.

\bibitem{barraud20207}
S.~Barraud, B.~Previtali, C.~Vizioz, J.~M. Hartmann, J.~Sturm, J.~Lassarre,
  C.~Perrot, P.~Rodriguez, V.~Loup, A.~Magalhaes-Lucas, R.~Kies, G.~Romano,
  M.~Cassé, N.~Bernier, A.~Jannaud, A.~Grenier, and F.~Andrieu,
  ``{{7-Levels-Stacked Nanosheet GAA Transistors for High Performance
  Computing}},'' in \emph{2020 IEEE Symposium on VLSI Technology}, Jun. 2020,
  pp. 1--2, {DOI} :
  {\href{https://doi.org/10.1109/VLSITechnology18217.2020.9265025}{10.1109/VLSITechnology18217.2020.9265025}}.

\bibitem{lee2004novel}
S.-Y. Lee, E.-J. Yoon, S.-M. Kim, C.~W. Oh, M.~Li, J.-D. Choi, K.-H. Yeo, M.-S.
  Kim, H.-J. Cho, S.-H. Kim, D.-W. Kim, D.~Park, and K.~Kim, ``{{A novel sub-50
  nm multi-bridge-channel MOSFET (MBCFET) with extremely high performance}},''
  in \emph{Digest of Technical Papers. 2004 Symposium on VLSI Technology,
  2004.}, Jun. 2004, pp. 200--201, {DOI} :
  {\href{https://doi.org/10.1109/VLSIT.2004.1345478}{10.1109/VLSIT.2004.1345478}}.

\bibitem{huang2020high}
X.~Huang, C.~Liu, Z.~Tang, S.~Zeng, L.~Liu, X.~Hou, H.~Chen, J.~Li, Y.-G.
  Jiang, D.~W. Zhang, and P.~Zhou, ``{{High Drive and Low Leakage Current MBC
  FET with Channel Thickness 1.2nm/0.6nm}},'' in \emph{2020 IEEE International
  Electron Devices Meeting (IEDM)}, Dec. 2020, pp. 12.1.1--12.1.4, {DOI} :
  {\href{https://doi.org/10.1109/IEDM13553.2020.9371941}{10.1109/IEDM13553.2020.9371941}}.

\bibitem{yoon2004sub}
E.-J. Yoon, S.-Y. Lee, S.-M. Kim, M.-S. Kim, S.~H. Kim, L.~Ming, S.~Suk,
  K.~Yeo, C.~W. Oh, J.~dong Choe, D.~Choi, D.-W. Kim, D.~Park, K.~Kim, and
  B.-I. Ryu, ``{{Sub 30 nm multi-bridge-channel MOSFET (MBCFET) with metal gate
  electrode for ultra high performance application}},'' in \emph{IEDM Technical
  Digest. IEEE International Electron Devices Meeting, 2004.}, Dec. 2004, pp.
  627--630, {DOI} :
  {\href{https://doi.org/10.1109/IEDM.2004.1419244}{10.1109/IEDM.2004.1419244}}.

\bibitem{ritzenthaler2018vertically}
R.~Ritzenthaler, H.~Mertens, V.~Pena, G.~Santoro, A.~Chasin, K.~Kenis,
  K.~Devriendt, G.~Mannaert, H.~Dekkers, A.~Dangol, Y.~Lin, S.~Sun, Z.~Chen,
  M.~Kim, J.~Machillot, J.~Mitard, N.~Yoshida, N.~Kim, D.~Mocuta, and
  N.~Horiguchi, ``{{Vertically Stacked Gate-All-Around Si Nanowire CMOS
  Transistors with Reduced Vertical Nanowires Separation, New Work Function
  Metal Gate Solutions, and DC/AC Performance Optimization}},'' in \emph{2018
  IEEE International Electron Devices Meeting (IEDM)}, Jun. 2018, pp.
  21.5.1--21.5.4, {DOI} :
  {\href{https://doi.org/10.1109/DRC.2018.8442137}{10.1109/DRC.2018.8442137}}.

\bibitem{mertens2016vertically}
H.~Mertens, R.~Ritzenthaler, A.~Chasin, T.~Schram, E.~Kunnen, A.~Hikavyy, L.-A.
  Ragnarsson, H.~Dekkers, T.~Hopf, K.~Wostyn, K.~Devriendt, S.~A. Chew, M.~S.
  Kim, Y.~Kikuchi, E.~Rosseel, G.~Mannaert, S.~Kubicek, S.~Demuynck, A.~Dangol,
  N.~Bosman, J.~Geypen, P.~Carolan, H.~Bender, K.~Barla, N.~Horiguchi, and
  D.~Mocuta, ``{{Vertically stacked gate-all-around Si nanowire CMOS
  transistors with dual work function metal gates}},'' in \emph{2016 IEEE
  International Electron Devices Meeting (IEDM)}, Dec. 2016, pp.
  19.7.1--19.7.4, {DOI} :
  {\href{https://doi.org/10.1109/IEDM.2016.7838456}{10.1109/IEDM.2016.7838456}}.

\bibitem{huang2021ultrathin}
X.~Huang, C.~Liu, S.~Zeng, Z.~Tang, S.~Wang, X.~Chen, D.~W. Zhang, and P.~Zhou,
  ``{{Ultrathin Multibridge Channel Transistor Enabled by van der Waals
  Assembly}},'' \emph{Advanced Materials}, vol.~33, no.~37, p. 2102201, Aug.
  2021, {DOI} :
  {\href{https://doi.org/10.1002/adma.202102201}{10.1002/adma.202102201}}.

\bibitem{liu2021two}
C.-J. Liu, Y.~Wan, L.-J. Li, C.-P. Lin, T.-H. Hou, Z.-Y. Huang, and V.~P.-H.
  Hu, ``{{2D Materials-Based Static Random-Access Memory}},'' \emph{Advanced
  Materials}, p. 2107894, Dec. 2021, {DOI} :
  {\href{https://doi.org/10.1002/adma.202107894}{10.1002/adma.202107894}}.

\bibitem{xia2022wafer}
Y.~Xia, L.~Zong, Y.~Pan, X.~Chen, L.~Zhou, Y.~Song, L.~Tong, X.~Guo, J.~Ma,
  S.~Gou, Z.~Xu, S.~Dai, D.~W. Zhang, P.~Zhou, Y.~Ye, and W.~Bao,
  ``{{Wafer-Scale Demonstration of MBC-FET and C-FET Arrays Based on
  Two-Dimensional Semiconductors}},'' \emph{Small}, vol.~18, no.~20, p.
  2107650, Apr. 2022, {DOI} :
  {\href{https://doi.org/10.1002/smll.202107650}{10.1002/smll.202107650}}.

\bibitem{zhou2018three}
R.~Zhou and J.~Appenzeller, ``{{Three-Dimensional Integration of Multi-Channel
  MoS\textsubscript{2} Devices for High Drive Current FETs}},'' in \emph{2018
  76th Device Research Conference (DRC)}, Jun. 2018, pp. 1--2, {DOI} :
  {\href{https://doi.org/10.1109/DRC.2018.8442137}{10.1109/DRC.2018.8442137}}.

\bibitem{xiong2021demonstration}
X.~Xiong, A.~Tong, X.~Wang, S.~Liu, X.~Li, R.~Huang, and Y.~Wu,
  ``{Demonstration of Vertically-stacked CVD Monolayer Channels:
  MoS\textsubscript{2} Nanosheets GAA-FET with I\textsubscript{ON} $>$ 700
  $\mu$A and MoS\textsubscript{2}/WSe\textsubscript{2} CFET},'' in \emph{2021
  IEEE International Electron Devices Meeting (IEDM)}, 2021, pp. 7.5.1--7.5.4,
  {DOI} :
  {\href{https://doi.org/10.1109/IEDM19574.2021.9720533}{10.1109/IEDM19574.2021.9720533}}.

\bibitem{majumdar2014scalability}
K.~Majumdar, C.~Hobbs, and P.~D. Kirsch, ``{{Benchmarking Transition Metal
  Dichalcogenide MOSFET in the Ultimate Physical Scaling Limit}},'' \emph{IEEE
  Electron Device Letters}, vol.~35, no.~3, pp. 402--404, Jan. 2014, {DOI} :
  {\href{https://doi.org/10.1109/LED.2014.2300013}{10.1109/LED.2014.2300013}}.

\bibitem{rahman2003theory}
A.~Rahman, J.~Guo, S.~Datta, and M.~Lundstrom, ``{{Theory of ballistic
  nanotransistors}},'' \emph{IEEE Transactions on Electron Devices}, vol.~50,
  no.~9, pp. 1853--1864, Aug. 2003, {DOI} :
  {\href{https://doi.org/10.1109/TED.2003.815366}{10.1109/TED.2003.815366}}.

\bibitem{Cheng20222Dinterfaces}
Z.~Cheng, H.~Zhang, S.~T. Le, H.~Abuzaid, G.~Li, L.~Cao, A.~V. Davydov, A.~D.
  Franklin, and C.~A. Richter, ``{Are 2D Interfaces Really Flat?}'' \emph{ACS
  Nano}, vol.~16, no.~4, pp. 5316--5324, Apr 2022, {DOI} :
  {\href{https://doi.org/10.1021/acsnano.1c11493}{10.1021/acsnano.1c11493}}.

\bibitem{murali2021accurate}
K.~Murali, M.~Dandu, K.~Watanabe, T.~Taniguchi, and K.~Majumdar, ``{{Accurate
  Extraction of Schottky Barrier Height and Universality of Fermi Level
  De-Pinning of van der Waals Contacts}},'' \emph{Advanced Functional
  Materials}, vol.~31, no.~18, p. 2010513, Feb. 2021, {DOI} :
  {\href{https://doi.org/10.1002/adfm.202010513}{10.1002/adfm.202010513}}.

\bibitem{roy2014field}
T.~Roy, M.~Tosun, J.~S. Kang, A.~B. Sachid, S.~B. Desai, M.~Hettick, C.~C. Hu,
  and A.~Javey, ``{{Field-effect transistors built from all two-dimensional
  material components}},'' \emph{ACS nano}, vol.~8, no.~6, pp. 6259--6264, Apr.
  2014, {DOI} : {\href{https://doi.org/10.1021/nn501723y}{10.1021/nn501723y}}.

\bibitem{Cheng2022HowTo}
Z.~Cheng, C.-S. Pang, P.~Wang, S.~T. Le, Y.~Wu, D.~Shahrjerdi, I.~Radu, M.~C.
  Lemme, L.-M. Peng, X.~Duan, Z.~Chen, J.~Appenzeller, S.~J. Koester, E.~Pop,
  A.~D. Franklin, and C.~A. Richter, ``{How to report and benchmark emerging
  field-effect transistors},'' \emph{Nature Electronics}, vol.~5, no.~7, pp.
  416--423, Jul 2022, {DOI} :
  {\href{https://doi.org/10.1038/s41928-022-00798-8}{10.1109/10.1038/s41928-022-00798-8}}.

\end{thebibliography}

\end{document}